# Disordered Microporous Sandia Octahedral Molecular Sieves are Tolerant to Neutron Radiation

*Rana Faryad Ali[†], Melanie Gascoine[†], Krzysztof Starosta[†], Byron D. Gates[†]\**

Department of Chemistry and 4D LABS, Simon Fraser University, 8888 University Drive, Burnaby, BC, V5A 1S6, Canada

\* E-mail: bgates@sfu.ca

**Abstract**

Materials that possess a porous and defected structure can have a range of useful properties that are sought after, which include their tolerance to nuclear radiation, ability to efficiently store and release isotopes, to immobilize nuclear waste, and to exhibit phase stability even at elevated temperatures. Since nanoscale pores and surface structures can serve as sinks for radiation-induced amorphization, one dimensional (1D) porous nanorods due to their high surface-to-volume ratio have the potential for use as advanced materials in nuclear science applications. In this study, we demonstrate a synthesis and a detailed analysis of microporous 1D octahedral molecular sieves of disodium diniobate hydrate [($Na_2Nb_2O_6 \cdot H_2O$) or Sandia Octahedral Molecular Sieves (SOMS)]. In addition, the stability of these SOMS is evaluated following their exposure to elevated temperatures and neutron irradiation. A surfactant-assisted solvothermal method is used to prepare these SOMS-based nanorods. This relatively low temperature, solution-phase approach can form crystalline nanorods of microporous $Na_2Nb_2O_6 \cdot H_2O$. These 1D structures had an average diameter of ~50 nm and lengths >1 µm. The nanorods adopted a defected microporous phase and matched the *C2/c* space group, which also exhibited a resistance to radiation induced amorphization. The dimensions, phase, and crystallinity of the SOMS-based nanorods after exposure to a high incident flux of neutrons were comparable to those of the as-synthesized products. The radiation tolerance of these microporous SOMS could be useful in the design of materials for nuclear reactors, resilient nuclear fuels, thermally resilient materials, high temperature catalysts, and durable materials for the handling and storage of radioactive waste.

**Introduction**

Materials that can withstand extreme environments (e.g., energetic radiation, intense light, corrosion, and either reduced or elevated temperatures) have received significant attention for their utility in interstellar space travel,[1] nuclear reactors,[2,3] and nuclear waste storage and transportation.[4] Resistance to degradation from neutron irradiation is of critical importance to the design of materials for the next-generation of nuclear power reactors, for improved methods of nuclear waste immobilization, and to enable interstellar space exploration.[5] The fundamental premise of a material's tolerance to radiation revolves around its ability to endure exposure to energetic radiation (e.g., neutrons, protons) while retaining many of their original properties, such as mechanical integrity.[6,7] The probability that a material will undergo a sufficient degree of damage such that they will lose their sought after properties will depend on the type, energy, and flux of the incident radiation, as well as the duration of exposure to this radiation.[8] The interactions between nuclear radiation and matter can be classified as processes resulting in (i) radiochemistry (i.e., ionization and free radical production) and (ii) atomic displacement collisions in crystalline solids.[9] In general, neutral radiation particles (e.g., neutrons) lose their energy through non-ionizing processes such as displacement reactions within materials. Such displacement reactions can knock-out atoms from the irradiated materials and/or change the arrangement of atoms within a crystalline lattice leading to amorphization of the materials.[11] On the other hand, processes resulting in material ionization are primarily associated with the interactions of incident charged particles (e.g., electrons, protons), which can result in structural transformations at the atomic scale. Materials that are potentially radiation tolerant will exhibit the ability to endure exposure to energetic radiation (e.g., neutrons) over a range of fluxes and under potentially extreme environmental conditions (e.g., lowered or elevated temperatures) for long periods of time. During this exposure, these materials will need to withstand radiation-induced structural transformations (e.g., amorphization, swelling, embrittlement).

A range of properties can be designed into new materials through the preparation of nanomaterials and/or nanostructured materials. The ability to manipulate the atomic-scale to microscale compositions and structures of materials using an array of techniques are enabling metallurgists and materials scientists to create a variety of customized materials.[8,10–14] Strategies that include the purposeful introduction of defects within crystal structures or increasing the surface area of materials have exhibited an improved tolerance to exposure to radiation including energetic neutrons. For example, different types of defects such as twinning boundaries (TB), grain boundaries (GB), and phase boundaries present in alloys, ceramics or complex oxides can improve the resistance of these materials towards prolonged exposure to energetic neutron irradiation.[2,15] Materials with a relatively large surface area, including porous structures, have also exhibited a great potential to resist exposure to radiation.[7,10,16] It has been demonstrated that defects present in porous materials are capable of radiation tolerance through material displacements at point defects and through an enhanced diffusion that alleviates swelling that would otherwise be caused by the accumulation of fission gas.[7,10,16] Additionally, the formation of defects within a porous materials can enhance its radiation tolerance through providing sinks for radiation-induced point defects, avoiding the accumulation of such induced defects due to the presence of a relatively high density of sinks provided by the interfaces therein.[7,10,16] It is, however, important that the defects incorporated into these porous materials have a sufficient mobility to migrate to the surfaces to provide such radiation tolerance, and that the particle morphology does not change drastically when subjected to energetic radiation.[10,16,17] One-dimensional (1D) materials such as porous nanorods with a high-surface-area-per unit volume could be ideal for use in applications that require exposure to extreme levels of neutron radiation.[10,16,17,15] These 1D materials may offer an enhanced ability to recover from radiation-induced damage through facilitating defect recovery through the proximity of surfaces therein through a high porosity and low-dimensional effects. Such features of these nanoscale, porous materials may enhance the diffusion of vacancies and induced interstitial defects therein in contrast to the same properties observed in bulk materials.

In this study, we prepared nanorods of a disordered microporous phase of disodium diniobate through a surfactant-assisted solvothermal method. This solvothermal method resulted in the formation of crystalline nanorods composed of disodium diniobate hydrate ($Na_2Nb_2O_6 \cdot H_2O$) otherwise known as Sandia Octahedral Molecular Sieves (SOMS). This synthesis was able to achieve a crystalline product without the need for additional calcination to induce crystallization. The nanorods contained a pure phase that possess a disordered microporous structure belonging to the *C2/c* space group. The presence of natural defects within the structure of these SOMS-based nanorods were also found to be resistant to radiation induced transformations (e.g., amorphization, changes in particle morphology). These 1D porous materials could be used as refractory materials for extreme temperature applications. In addition, the demonstrated solution-phase synthesis could be utilized in the future to prepare nanomaterials of other types of octahedral molecular sieves, pyrochlores, and other perovskite phases.

**Materials and Supplies**

All of the chemicals used herein were of an analytical grade and were used as received without further purification. Niobium pentachloride ($NbCl_5$, 99.0 %) was obtained from Sigma-Aldrich. The citric acid ($C_6H_8O_7$, 99.0 %) and sodium hydroxide (NaOH, 99.9 %) were purchased from Caledon Laboratories Ltd. and VWR, respectively. Glycerol ($C_3H_8O_3$, 99.0 %) was purchased from Caledon Laboratories Ltd., and anhydrous ethanol was obtained from Commercial Alcohols.

**Synthesis of Sandia Octahedral Molecular Sieves (SOMS)**

A 0.2 M aqueous solution of $NbCl_5$ was prepared by dissolving 0.216 g of $NbCl_5$ in 4 mL $H_2O$. Into this solution, 2 mL of 0.60 M aqueous solution of the citric acid monohydrate and 2 mL of a 0.50 M aqueous solution of sodium hydroxide (NaOH) was added, which resulted in the formation of white precipitates. The mixture was heated at 200 °C for 12 h in a Teflon lined autoclave (Model No. 4749, Parr Instruments Co., Moline, IL USA). After cooling, the precipitates were isolated via centrifugation (8,000 rpm for 10 min) and washed by dispersion in 10 mL of water. The obtained precipitates were dried at 80 °C to yield the precursor materials. A 0.5 M aqueous solution of NaOH (9.0 mL) and 3 mL of glycerol were mixed to form a solution into which 40.0 mg of the precursor material was added and the final mixture stirred for 1 h at room temperature. The resulting suspension was transferred to a 23 mL Teflon lined autoclave, which was sealed and heated at 200 ºC for 4 h. After cooling to room temperature, white precipitates were isolated from the solution via a process of centrifugation (AccuSpin Model No. 400, Fisher Scientific) at 8,000 rpm for 20 min and decanting of the supernatant. The isolated solids were washed three times by re-suspending in 10 mL of ethanol and repeating the process of centrifugation and decanting of the solution. The purification process was repeated three more times with 10 mL of deionized water (18 MΩ·cm, produced using a Barnstead NANOpure DIamond water filtration system). The purified product was dried at 70 ºC for 12 h to remove residual water prior to further characterization.

**Characterization of the Products**

The morphology, dimensions, and crystallinity of the SOMS-based nanorods were characterized using an FEI Osiris X-FEG scanning/transmission electron microscope (S/TEM) operated at an accelerating voltage of 200 kV. Analyses by energy dispersive X-ray spectroscopy (EDS) were performed using the same S/TEM system, which was equipped with a Super-X EDS system using ChemiSTEM Technology integrating the signal from four spectrometers. Samples for TEM analysis were prepared by dispersing the purified products in ethanol followed by drop-casting 5 µL of each suspension onto separate TEM grids (300 mesh copper grids coated with formvar/carbon) purchased from Cedarlane Labs. Each TEM grid was dried under vacuum at ~230 Torr for at least 20 min prior to analysis. The TEM aperture used to measure the selected area electron diffraction (SAED) from multiple nanoparticles was 40 µm. The source of the copper (Cu) signals in the EDS spectra were from the Cu TEM grids used to hold the samples.

Crystallinity and phase of the product were further assessed by X-ray diffraction (XRD). The XRD patterns were acquired with a Rigaku R-Axis Rapid diffractometer equipped with a 3-kW sealed copper tube source (Kα radiation, λ = 0.15418 nm) collimated to 0.5 mm. Powder samples were packed into cylindrical recesses drilled into glass microscope slides (Leica 1 mm Surgipath Snowcoat X-tra Micro Slides) for acquiring XRD patterns of the products.

Purity and phase of the product were further characterized using Raman spectroscopy techniques. Raman spectra were collected using a Renishaw inVia Raman microscope with a 50x LWD objective lens (Leica, 0.5 NA), and a 514 nm laser (argon-ion laser, Model No. Stellar-Pro 514/50) set to 100 % laser power with an exposure time of 30 s. The Raman spectrometer was calibrated by collecting the spectrum of a polished silicon (Si) standard with a distinct peak centered at 520 cm$^{-1}$. The Raman spectra for the samples were acquired from 100 to 1,000 cm$^{-1}$ using a grating with 1,800 lines/mm and a scan rate of 10 cm$^{-1}$ per second.

The thermal stability of the products was characterized by thermogravimetric analyses (TGA) performed using a SHIMADZU TGA-50 thermogravimetric analyzer. Dried powder samples were held in platinum sample pans and heated from 30 to 850 °C at a rate of 1 °C min$^{-1}$ under a nitrogen gas atmosphere.

The stability of the SOMS when exposed to radiation was assessed by irradiating the sample with high energy neutrons. Neutrons with an energy of 14.1 MeV were produced using a Thermo-Fisher P385 deuterium/tritium generator operated with a terminal voltage of 130 kV and at a current of 69.5 µA. The samples were held 6 cm away from the center of the neutron production target within the neutron generator. The samples were irradiated for 72 h with ∼0.21 × 10$^9$ (0.21 billion) neutrons per second emitted into the full solid angle of 4π (i.e., solid angle of a sphere). The calculated fluence was ∼1.20 × 10$^{11}$ neutrons per cm$^2$ at the sample during the 72 h exposure time. The cross-sectional area of the solid sample was 1.31 cm$^2$, which yielded a total exposure of the sample to ∼1.57 × 10$^{11}$ neutrons.

**Results and Discussion**

A disordered form of microporous $Na_2Nb_2O_6 \cdot H_2O$ or SOMS-based nanorods were prepared using a hydrothermal synthesis (**Figure 1**). This method for making SOMS-based materials introduced the use of a hydrothermal process, which did not involve the use of air-sensitive or expensive starting materials (e.g., metal alkoxides). In comparison to the traditional method of preparing SOMS that require >3 days of synthesis followed by heat treatment >200 °C, our approach yielded single crystalline SOMS at much shorter reaction times and lower temperatures.[18–21] In an effort to make SOMS more attractive to industrial scale-up and commercialization, we used niobium chloride (NbCl$_5$) and sodium hydroxide (NaOH) as precursors due to their ease of access and relatively low cost. The niobium pentachloride precursor was dissolved in an aqueous 0.50 M solution of NaOH and an aqueous 0.60 M solution of citric acid monohydrate at ambient conditions. The niobium precursor underwent a controlled hydrolysis in the presence of the citrate ions that resulted in the formation of white precipitates. This suspension was subsequently subject to a hydrothermal treatment in a Teflon lined autoclave at 200 °C for 12 h. This step produced a precursor that was used for the preparation of single crystalline SOMS. We characterized the morphology and crystallinity of the precursor by transmission electron microscopy (TEM) techniques. An aggregated, random network of nanoparticles was observed in the hydrolyzed precursor, which indicated an onset of the condensation process in this material (**Figure S1**). Crystallinity and phase of this sample were characterized by selected area electron diffraction (SAED) and high-resolution TEM (HRTEM) techniques (**Figure S1**). We did not observe any significant diffraction rings or spot patterns in the SAED analyses, which indicated an amorphous or partially crystalline nature of the sample. The HRTEM analyses also revealed the presence of lattice fringes in a few particles of this sample that further confirmed the formation of both amorphous and semi-crystalline components. The semi-crystalline nature of the sample was further revealed by the relatively broad peaks observed by powder X-ray diffraction (XRD) analyses and in the associated Raman spectroscopy bands (**Figures S2, S3**). Composition and elemental distribution within the solids of the obtained precursor were analyzed by energy dispersed X-ray spectroscopy (EDS). Dark-field based TEM images were correlated with EDS maps, enabling a correlation of composition therein to the morphology of the observed nanoparticles. The EDS maps revealed a relatively uniform distribution of Na, Nb, and O within the nanoparticles (**Figure S4**). This precursor was subsequently treated by a hydrothermal process in a Teflon lined autoclave at 200 °C for 4 h in the presence of glycerol as a surfactant and an aqueous solution of 0.5 M NaOH to yield single-crystalline SOMS. We devised a two-step synthesis to prepare the SOMS product to avoid the introduction of structural and phase impurities that can include an "unidentified hexaniobate intermediate" phase, and this approach combines all the necessary framework elements prior to the formation of the SOMS framework.[18–21] The referred to hexaniobate phase has been found to be stable and is able to resist the transformation into SOMS for at least 99 h when pursued in a single-step hydrothermal processes. Through the method introduced herein, during the second step of the synthesis of the SOMS a hydrothermal treatment is used to crystallize the precursor into a microporous form of SOMS without the inclusion of hexaniobate impurities. It was found that washing of precursor with water was an important step to avoid this impurity.

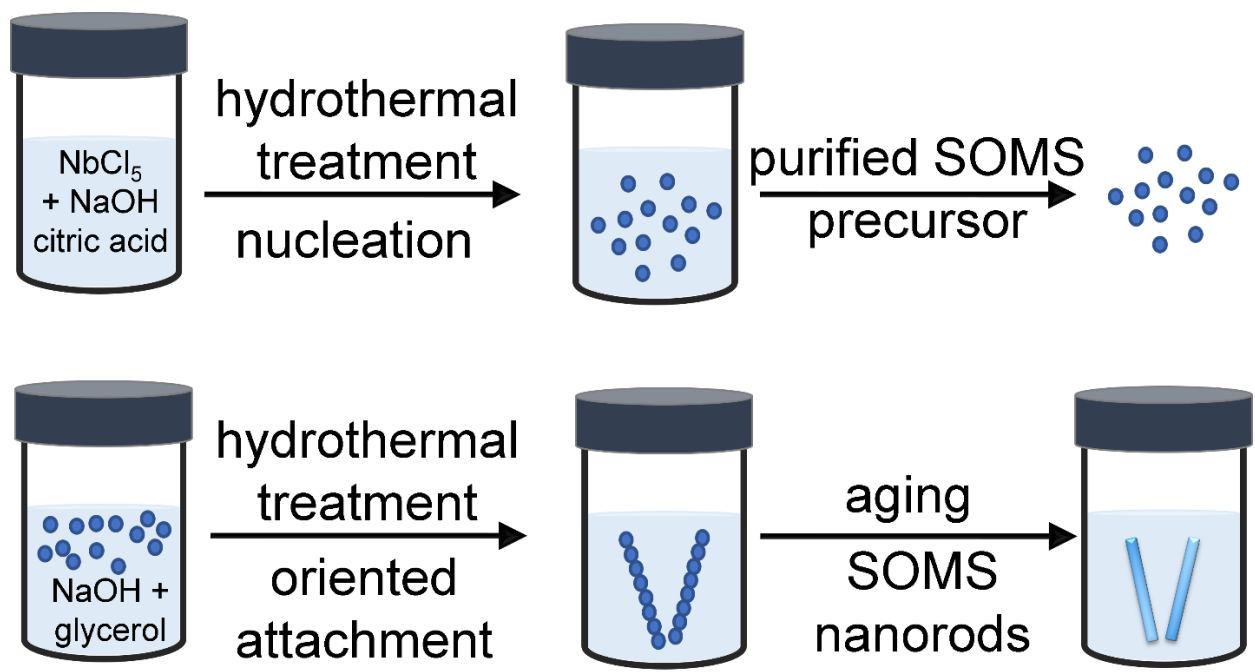

**Figure 1.** A mechanism proposed for the synthesis of SOMS-based nanorods from a purified precursor. During the hydrothermal treatment of this precursor, seeds are nucleated and grow in the presence of glycerol incorporated as a surfactant. The resulting surfactant capped nanocrystals underwent a process of oriented attachment to form one-dimensional nanostructures and a subsequent aging process resulted in the formation of SOMS-based nanorods.

The synthesis of SOMS-based nanorods from their precursor was achieved using a hydrothermal treatment. The 1D product formed through a process of crystallization and oriented attachment. The amorphous precursor to SOMS was added in an aqueous solution of NaOH (0.5 M) containing glycerol, which was added to control the growth of the SOMS nanorods. The resulting mixture was subjected to hydrothermal treatment wherein NaOH provided the hydroxyl ions necessary for adjusting the surface energy and to promote growth of the nanoparticles and the hydrothermal treatment required to produce the oxidizing conditions for the reaction. The growth of anisotropic, SOMS-based nanorods from the precursor material and the nanoparticles formed therein likely occurred through a process of oriented attachment.[22,23] This oriented attachment is likely driven by a permanent dipole moment within the crystallites of the precursor material and subsequently in the nanoparticles formed in the hydrothermal synthesis due to an anisotropic distribution of surface charges. A dipole arises from differences in the atomic distribution of Nb and O on the surfaces of these precursors to the nanorods and from differences in the electronegativities of the elements within these materials.[22,23] A relatively large dipole moment is predicted to form along the [001] direction of SOMS-based materials. Smaller dipole moments can form along other crystal directions but were anticipated to have less of an influence on the self-organization of the nanoparticles. The adsorption of surfactant molecules onto the surfaces of the forming nanoparticles may have also significantly decreased the surface energies of some facets in comparison to the {001} facets.[22,23] These combined effects led to a relatively large dipole moment along the [001] direction, and the interactions of the dipoles between individual nanoparticles resulted in an oriented attachment of these nanocrystals along the [001] direction. The result was, vide infra, the formation of 1D nanorods.

Crystallinity, phase, and the purity of each in the as-obtained products were analyzed using powder X-ray diffraction (XRD) techniques (**Figure 2**). The XRD patterns suggested that the product contained a pure phase of SOMS, with a formula of $Na_2Nb_2O_6 \cdot H_2O$ in a monoclinic lattice assigned to a *C2/c* space group with lattice spacings of *a* = 17.05 Å, *b* = 5.03 Å, *c* = 16.49 Å. The sample appears to have a relatively high degree of crystallization as observed in the XRD results. A semi-indexed XRD plot indicated the matching of major XRD reflections with the reported ICSD No. 55415 **(Figure S5).** The crystal structure of the hydrothermal product contains $[NbO_6]$ and $[NaO_6]$ octahedra with the remaining Na species occupying the channel sites within the crystalline framework. The $[NbO_6]$ octahedra were connected in a manner that formed double chains that ran parallel to [010] direction, whereas the $[NaO_6]$ octahedral were connected in layers that were parallel to (001). The octahedral formed via edge sharing. The layers of $[NaO_6]$ octahedral alternated with those layers containing the double chains of $[NbO_6]$ octahedral along the length of the *c*-axis. The assembly of these alternating layers formed a three-dimensional network of atomic-scale channels. Both the $[NbO_6]$ and $[NaO_6]$ octahedra were irregular due to edge sharing (in contrast to corner sharing that commonly occurs in framework structures). This edge sharing of $[NbO_6]$ or $[NaO_6]$ octahedra also minimized the mismatch in the crystalline lattices between the $[NbO_6]$ and $[NaO_6]$ layers. In addition, the $[NaO_6]$ octahedra were presumably more flexible than the $[NbO_6]$ octahedra, which resulted in larger deviations of the O−Na−O bond angles than for the O-Nb-O bond angles from 90° or 180° that would otherwise have been observed in perfect octahedral molecular geometries. For example, the smallest O−Na−O bond angle was 64.3°, whereas the smallest O−Nb−O bond angle was 74.8°. The additional incorporation of Na into the lattice resides along one-dimensional channels that form parallel to the *b*-axis and each of these Na species are coordinated to four O atoms in a distorted, square-planar geometry. Another

O is 2.84 Å from Na(3) and can be considered to be bonded to Na(3). More specifically, Na(3) occupies one of the two sites displaced from the center of a rectangle, rather than at the center itself. Thus, each Na(3) site has an occupancy of 50%. This structure is unique in the sense that Na, typically an extra cation in the framework, also participates in the formation of the framework leading to the creation of distorted materials containing defect sites in the lattice. Similar structural features and defect sites within the lattice have been observed in some lithosilicate zeolites in which Li cations can occur both within the framework and as extra framework species, though in that case the framework consists of tetrahedral species.

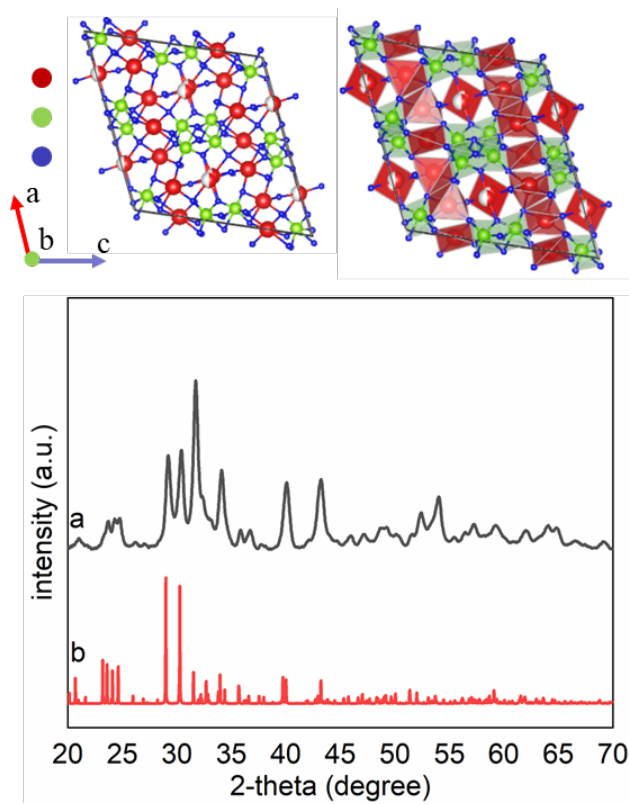

**Figure 2.** The crystal structure of Sandia octahedral molecular sieves ($Na_2Nb_2O_6 \cdot H_2O$ or SOMS) constructed and visualized using VESTA. The crystallographic parameters were obtained from the CIF file corresponded to ICSD No. 55415. Powder X-ray diffraction (XRD) patterns are plotted for: (a) SOMS prepared by a solvothermal synthesis; and (b) a reported reference sample of $Na_2Nb_2O_6 \cdot H_2O$ (ICSD No. 55415). The major reflections associated with the products matched those of the reported reference sample.

Analyses performed by transmission electron microscopy indicated the formation of nanorods prepared from single-crystalline SOMS with relatively uniform diameters (**Figure 3**). The average length of the nanorods was above 1 μm and their widths ranged from 30 to 50 nm. Crystallinity and phase of the nanorods were further investigated using electron diffraction techniques. A well-defined spot pattern obtained by SAED was observed for individual nanorods, which confirmed the formation of single-crystalline products (**Figure 3c**). The crystallinity of these nanoparticles was further evaluated by HRTEM (**Figure 3d**). The periodic fringe patterns observed by HRTEM for some of these nanoparticles had a *d*-spacing of 7.8 Å along their length. This inter-planar spacing matched the spacing of the (200) planes of the *monoclinic* unit cell of SOMS, which implies a one-dimensional growth of nanorods along the [100] direction that arose from a process of oriented attachment. Composition and elemental distribution within the SOMS-based nanorods were analyzed by EDS techniques. Representative dark-field imaging and EDS-based elemental maps of the nanorods are shown in **Figure 4**. The elemental mapping reveals a uniform distribution of Na, Nb, and O throughout the nanorods. Quantitative analyses by EDS of these nanomaterials indicated the presence of Na, Nb and O in a mole ratio of ~1 : 1 : 3, which agreed with the theoretical molar ratios anticipated for $Na_2Nb_2O_6$.

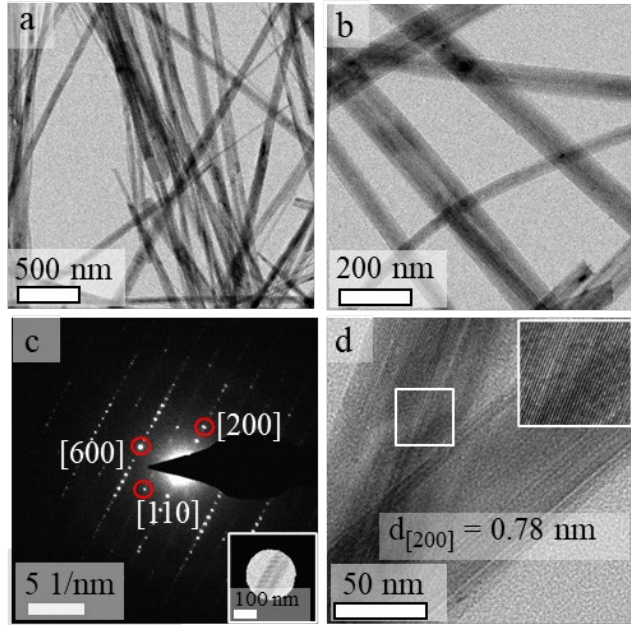

**Figure 3.** Nanorods of SOMS were characterized by: (a,b) transmission electron microscopy (TEM) imaging of multiple nanorods; (c) selected area electron diffraction (SAED) obtained from a single nanorod; and (d) high-resolution (HR) TEM image of a couple of nanorods. The inset image in (d) more clearly depicts the uniform lattice fringes within a single nanorod from the HRTEM image.

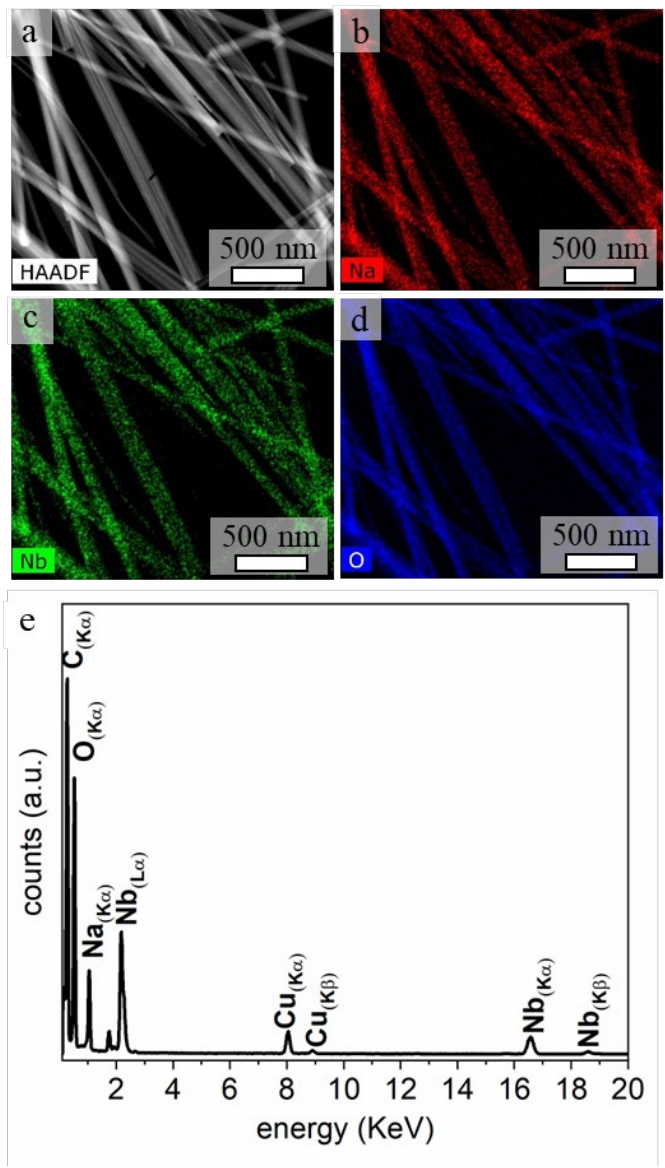

**Figure 4.** (a) High-angle annular dark-field (HAADF) scanning TEM (STEM) image of SOMS-based nanorods and corresponding elemental maps obtained by energy dispersed X-ray spectroscopy (EDS) for (b) Na, (c) Nb, and (d) O. (e) An EDS spectrum corresponding to the average composition of the nanorods, which further confirms the presence of Na, Nb, and O in the product. The source of the Cu signals in the spectrum was the TEM grid supporting the sample.

Composition, purity, and crystallinity of the SOMS-based nanorods were further characterized by Raman spectroscopy techniques (**Figure 5**). Raman spectroscopy can be used to differentiate between disordered and ordered structures of materials. In the monoclinic *C2/c* structure, all atoms occupy non-centrosymmetric sites, and their vibrations are Raman active. Characteristic Raman bands at ~210 cm$^{-1}$, ~374 cm$^{-1}$, ~460 cm$^{-1}$, ~638 cm$^{-1}$ and ~884 cm$^{-1}$ observed from the analysis of the SOMS nanorods indicated the presence of an axially distorted octahedral framework within these materials.[21,24,25] On the basis of mass and bond length considerations, most of these bands can be assigned to modes involving mainly vibrations of oxygen atoms within the SOMS lattice. The stretching-type Raman modes are observed at higher frequencies than the bending and rotational Raman modes of the same species. The Raman band observed with the highest frequency (884 cm$^{-1}$) corresponds to a stretching vibration of the shortest bonds of Nb−O ≈ 1.8 Å, i.e., associated with the internal stretching vibrations of the NbO$_6$ octahedra. In ordered materials, the conservation of momentum results in only the zone center (Γ-point) phonons being Raman-allowed.[21,24,25] In disordered materials, however, the loss of translational symmetry results in an activation of all phonon modes (or all modes related to the disordered sublattice). The disorder in the crystal structure of SOMS can be attributed to the existence of corner and edge shared NbO$_6$ octahedra, which adopt a distorted structure due to the smaller size of Nb$^{5+}$ leading to these species not obeying Pauling's electrostatic valence rule.[26] This distortion in the octahedra within the SOMS framework indicate an anionic disorder due to an increased lattice strain that results from the substitution and mixing of the mismatched cations within the lattice. This analysis provides further evidence for the defects within and distortion to the structure within the SOMS-based nanorods.

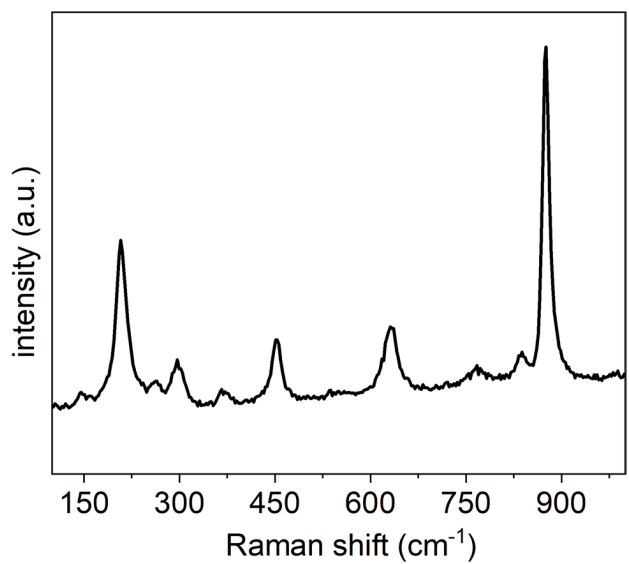

**Figure 5**. Room temperature Raman spectrum for the SOMS-based nanorods obtained using an excitation at 514 nm and a scan rate of 10 cm$^{-1}$ per second.

In addition, thermogravimetric analysis (TGA) was used to provide information on the thermal stability of the SOMS and a further confirmation of their composition (**Figure S6**). The SOMS exhibited two weight loss events upon treatment to elevated temperatures. A weight loss observed between 150 and 250 °C was associated with the loss of water and hydroxyls within the framework, which agrees with the reported variation from 7 to 8.5 wt% for previously synthesized SOMS of a range of compositions.[20] Below 150 °C the weight change is likely surface bound water, which may vary in its quantity depending on the methods used for sample preparation and variations in surface area within a particular sample. No further weight loss was observed up to 900 °C indicating a relatively high degree of thermal stability apart from the earlier losses of water.

Molecular sieves have been shown to exhibit promising properties to assist in the clean-up and storage of radioactive waste, such as that produced from nuclear reactors.[27–29] The crystalline and porous framework of inorganic SOMS enable them to withstand the extremely caustic environments present for many nuclear isotopes. This property inspired us to explore the ability of SOMS to tolerate exposure to energetic neutrons. Radiation tolerance is measured as the ability of a material to resist undesirable radiation-induced phenomena such as swelling, amorphization, clustering of point defects, and formation of new crystalline phases.[2,5,15,30] These transformations often lead to significant volume changes in a material, to the formation of microcracks therein, and ultimately to the failure of both the structural integrity and properties of the material under irradiation. The radiation tolerance of materials can be predicted by the ability of a material to accommodate lattice point defects. Fast neutrons (e.g., neutrons with energies >1 MeV) passing through a material can result in the displacement of atoms therein, which can lead to structural damage (e.g., defects, vacancies, amorphization) and the temporary elevation of temperature within localized regions in the sample.[2,5,15,30] SOMS possess a disordered, porous structure that may make them promising candidates to resist the detrimental effects of radiation damage.

Another unique feature of SOMS is their ability to form nanorods. Mechanisms for radiation-induced damage mechanisms in 1D materials and porous (e.g., nanoporous and microporous) structures have been demonstrated to exhibit a dependence on particle size and crystalline lattice.[31,32] The primary mechanism of radiation induced damage in 1D materials with diameters <10 nm is through sputtering of atoms. One-dimensional materials with such small diameters are relatively less tolerant to radiation when increasing levels of damage are primarily associated with the formation of surface pits in the material due to consecutive strikes from incident radiation. These defects can form throughout the surfaces of these materials. Nanorods with larger diameters (*e.g.,* >10 nm) can also experience sputtering during irradiation, but due to its smaller surface to volume ratio, surface sputtering is a less dominant mechanism for structural damage.[31,32] For such 1D nanostructures with larger dimensions the dominant mechanism for radiation induced damage is the formation and accumulation of point defects, dislocation loops, amorphization, and clusters of defects. Contrary to non-porous 1D materials, those 1D structures that are porous can contain ligaments of different dimensions including nodes connecting different ligaments.[31,32] As a result of this design, the ability of 1D porous structures to tolerate radiation induced damage is more diverse than that of non-porous 1D materials. The nodes and ligaments of the porous materials provide additional mechanisms for the accumulation of damage, which improves the radiation tolerance in porous 1D structures. The formation of SOMS as nanorods with diameters >10 nm could exhibit a high degree of radiation tolerance.

The SOMS-based nanorods synthesized herein were studied for the stability of their phase, crystal structure, and composition following exposure to a flux of high energy neutrons. These nanorods were irradiated for 72 h by fast neutrons with an average energy of 14.1 MeV. The incident radiation had a fluence of ~$1.20 \times 10^{11}$ neutrons per $cm^2$. The nanorods were exposed to a total of ~$1.57 \times 10^{11}$ neutrons. The average size and shape of the nanorods after exposure to the incident neutrons were comparable to the nanorods before exposure to this radiation (**Figure 6**). No significant changes to the dimensions and shape of the nanorods after neutron irradiation indicates the stability of the microporous frameworks of the SOMS. Crystallinity of the irradiated nanorods was assessed by SAED and HRTEM. These analyses indicated that the nanorods remained crystalline after exposure to the fast neutrons (Figure 6c and d), but the less well-defined HRTEM images of the nanorods can be attributed to some amorphization in the product. The crystallinity and phase of these materials before and after the radiation treatment were also comparable, which indicated that the nanoparticles were tolerant to neutron radiation. The potential for atomic-scale changes, including the crystallinity of these nanomaterials was further characterized using aberration corrected TEM. This detailed analysis was used to compare the non-irradiated SOMS to the samples after irradiation (**Figure S7**). This analysis was performed while holding the samples at cryogenic temperatures to minimize potential damage from the incident, focused electron beam. These SAED and HRTEM analyses performed by aberration corrected electron microscopy indicated the single-crystalline nature of the SOMS-based nanorods was preserved following irradiation.

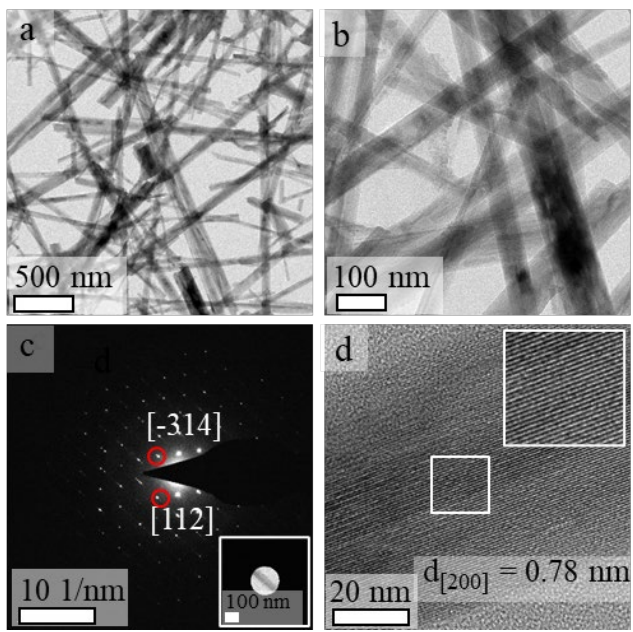

**Figure 6**. Characterization of the SOMS-based nanorods after their exposure to a flux of 14.1 MeV neutrons for 72 h included analyses by: (a,b) TEM, depicting multiple nanorods; (c) SAED of an individual nanorod; and (d) HRTEM of an individual nanorod.

An analysis of the nanoparticles by EDS after neutron irradiation demonstrated that the product had a similar composition to the as-synthesized $Na_2Nb_2O_6 \cdot H_2O$ nanorods (**Figure S8**). The nanorods exposed to neutrons were also characterized by XRD techniques. The resulting XRD patterns indicated that the phase of these nanorods matched that of the original product and the crystallinity of each were nearly identical (**Figure 7**). A decrease in the relative intensity of some of the reflections in the XRD plots for the nanorods following their exposure to fast neutrons were attributed to the presence of some amorphous regions and/or fragmented species in the otherwise crystalline product. No new peaks were observed in the diffraction patterns, indicating an absence of a phase transformation in the product after the exposure to radiation. Neutron induced structural changes in addition to bond fragmentation were not apparent in the crystallographic analyses of the nanorods. The potential for irradiation-induced structural changes and crystallographic transformations within the SOMS nanorods were further assessed by Raman spectroscopy (**Figure S9**). There was no shift in the Raman bands and new Raman bands were apparent in the samples following their neutron irradiation, indicating a high degree of ability of the SOMS-based nanorods to resist neutron irradiation induced transformations. Disordered microporous SOMS have a high degree of radiation tolerance due to their ability to accommodate lattice disorder and to recover from point defects. The disorder that provides this protection from incident radiation include the presence of anion vacancies and cation mixing within the crystalline lattice. These radiation tolerant materials could be explored in future studies for their ability to serve as materials for use in fusion or fission reactors, in materials for nuclear waste storage, or as components of radiation sensors.

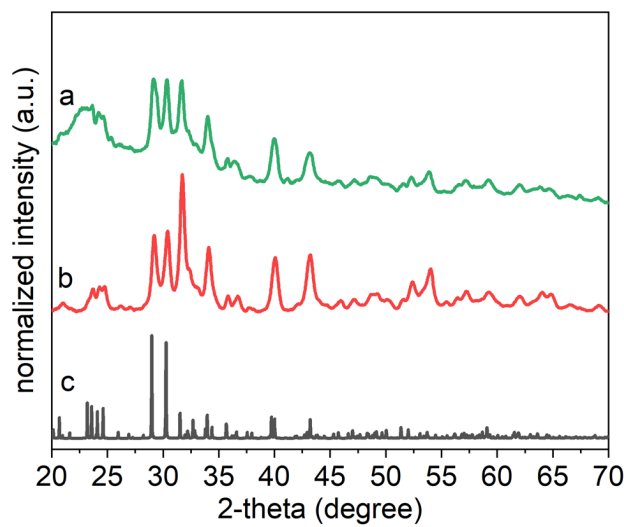

**Figure 7.** (a) Powder XRD patterns of SOMS after exposure to 14.1 MeV neutrons for 72 h. For comparison, powder XRD patterns are also included for (b) pristine SOMS before exposure to neutrons, and (c) a reported SOMS reference sample (ICSD No. 55415).

**Conclusions**

In summary, we introduced a solvothermal method to prepare relatively uniform, crystalline SOMS-based nanorods containing a microporous structure with atomic-scale defects within the lattice. These 1D nanostructures had average lengths >1 µm and diameters >50 nm. The solution-phase synthesis of these nanorods was carried out at a relatively low temperature (*e.g.*, 200 °C) in an aqueous environment over a reaction time of 16 h. Phase, purity, and composition of the product were characterized by integrating the results of XRD, Raman spectroscopy, TEM, and EDS analyses. An XRD analysis of the as-synthesized product indicated the formation of a pure microporous phase of $Na_2Nb_2O_6 \cdot H_2O$ having a defect containing structure associated with the *C2/c* space group. Phase and stability of these nanorods was evaluated by XRD, and TGA analyses indicated, following an initial loss of water, that these microporous SOMS were thermally stable up to ~900 °C. Stability of the SOMS following their exposure to fast neutrons was also studied by irradiating the sample with 14.1 MeV neutrons for 72 h for a total exposure of $\sim 1.20 \times 10^{11}$ neutrons per $cm^2$. After exposure to neutrons, no significant changes were observed to the size, shape, phase, and crystallinity of the nanorods. The stability of the SOMS-based nanorods to neutron radiation could be of particular interest for use as preparing radiation tolerant materials for use in applications requiring radiation-resistant materials that include windows for fusion reactors and containers for nuclear waste. This comprehensive study presents a new method to prepare radiation tolerant materials, which could be extended to prepare other types of radiation tolerant materials.

**Acknowledgements**

This work was supported in part by the Natural Sciences and Engineering Research Council (NSERC) of Canada (Discovery Grant No. RGPIN-2020-06522), and CMC Microsystems (MNT Grant No. 6719). This work made use of 4D LABS (www.4dlabs.com) and the Center for Soft

Materials shared facilities supported by the Canada Foundation for Innovation (CFI), British Columbia Knowledge Development Fund (BCKDF), Western Economic Diversification Canada, and Simon Fraser University. We also thank Dr. Jonathan Williams and Andrew Redey for their assistance to monitor the neutron radiation experiments.

**Conflicts of Interest**

There are no conflicts to declare.

# Graphical Abstract

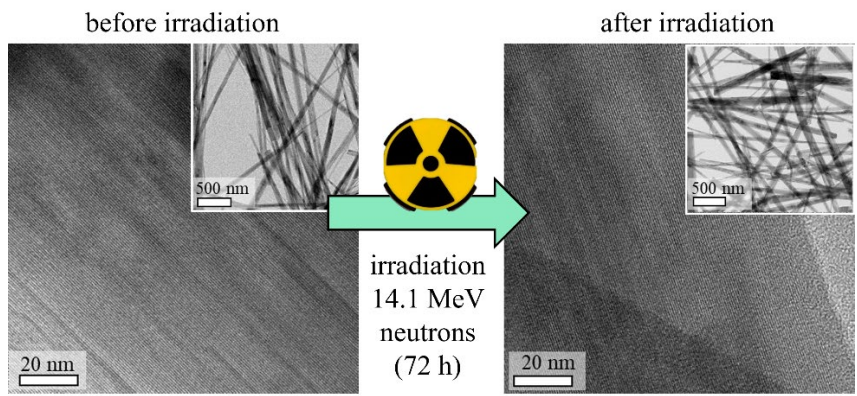

Nanorods were synthesized from disordered Sandia Octahedral Molecular Sieves (SOMS), and these materials characterized for their tolerance to neutron irradiation.

**Electronic Supplementary Information**

# Disordered Microporous Sandia Octahedral Molecular Sieves are Tolerant to Neutron Radiation

*Rana Faryad Ali†, Melanie Gascoine†, Krzysztof Starosta†, Byron D. Gates†\**

Department of Chemistry and 4D LABS, Simon Fraser University, 8888 University Drive, Burnaby, BC, V5A 1S6, Canada

\* E-mail: bgates@sfu.ca

<>
This work was supported in part by the Natural Sciences and Engineering Research Council (NSERC) of Canada (Discovery Grant No. RGPIN-2020-06522), and CMC Microsystems (MNT Grant No. 6719). This work made use of 4D LABS (www.4dlabs.com) and the Center for Soft Materials shared facilities supported by the Canada Foundation for Innovation (CFI), British Columbia Knowledge Development Fund (BCKDF), Western Economic Diversification Canada, and Simon Fraser University. We also thank Dr. Jonathan Williams and Andrew Redey for their assistance to monitor the neutron radiation experiments.

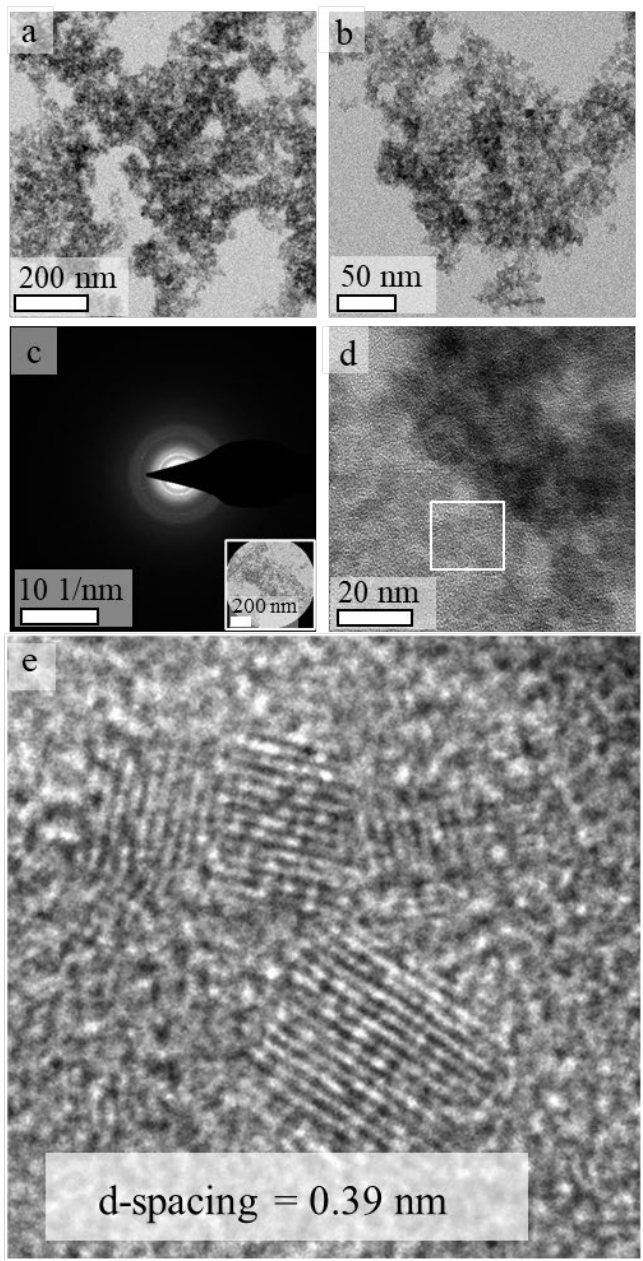

**Figure S1.** The precursor materials used to prepare Sandia octahedral molecular sieves (SOMS) as characterized by: (a, b) bright field transmission electron microscopy (TEM); (c) selected area electron diffraction (SAED); and (d) high resolution (HR) TEM. A magnified view of the region in (d) indicated by the white-box is shown in (e). This semi-crystalline product had a d-spacing of 0.39 nm as observed in some the nanoparticles by HRTEM analyses.

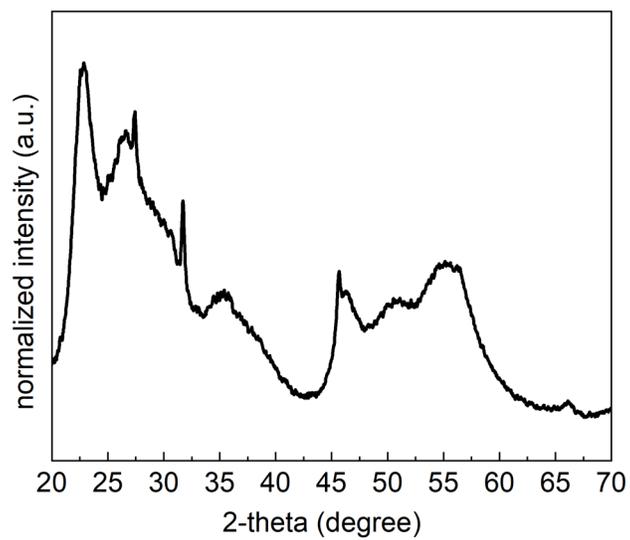

**Figure S2.** Powder X-ray diffraction (XRD) patterns associated with the precursor used to synthesize nanorods of SOMS through a solution-phase synthesis.

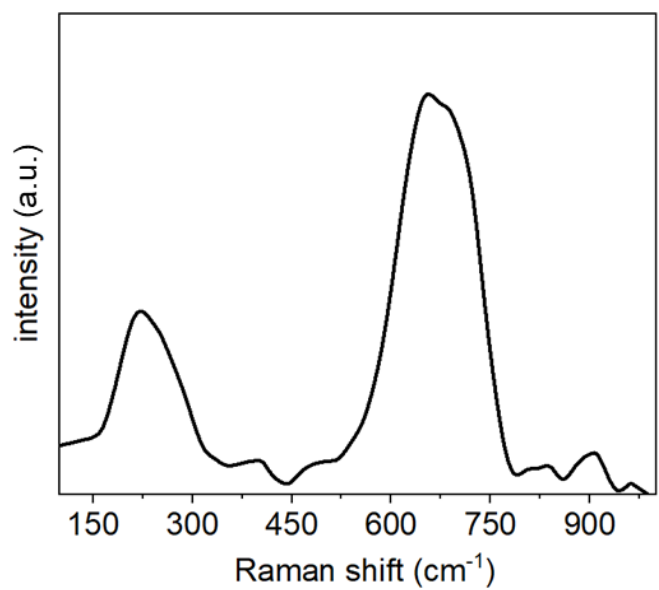

**Figure S3.** Room temperature Raman spectrum for the precursor material used to prepare the SOMS.

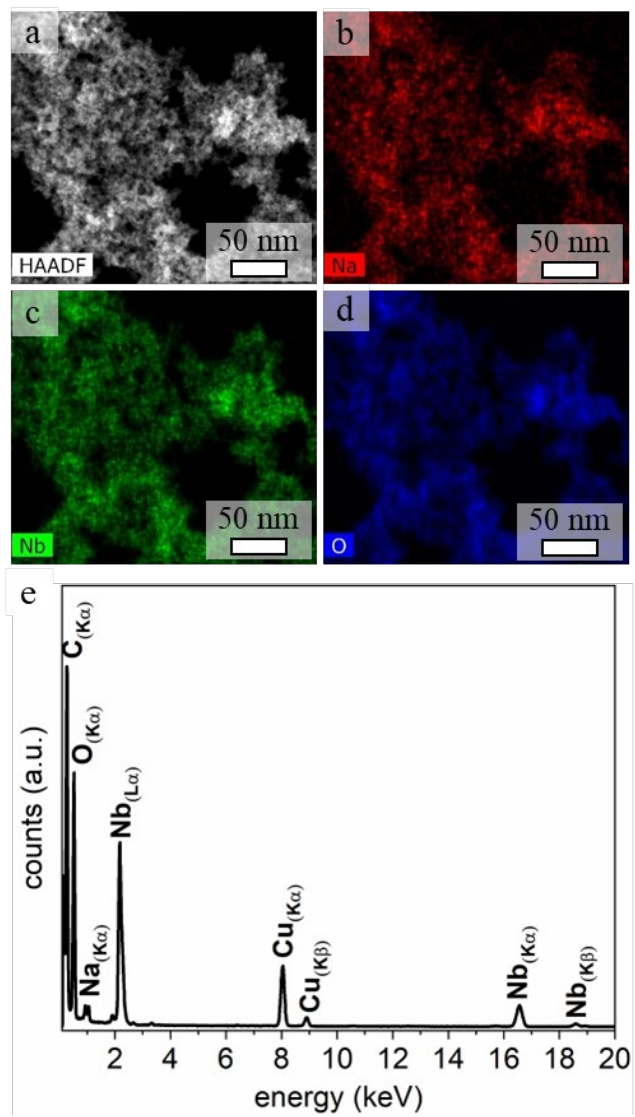

**Figure S4.** (a) High-angle annular dark-field (HAADF) scanning TEM (STEM) image of the nanoparticles present within the precursor to the SOMS, and corresponding maps of the elements within these materials as obtained by energy dispersed X-ray spectroscopy (EDS) for (b) Na, (c) Nb, and (d) O. (e) An EDS spectrum depicting the average spectral response of these nanomaterials, which further confirms the presence of Na, Nb, and O in this precursor material. The source of the Cu signals in the spectrum was the TEM grid used to support the sample for these analyses.

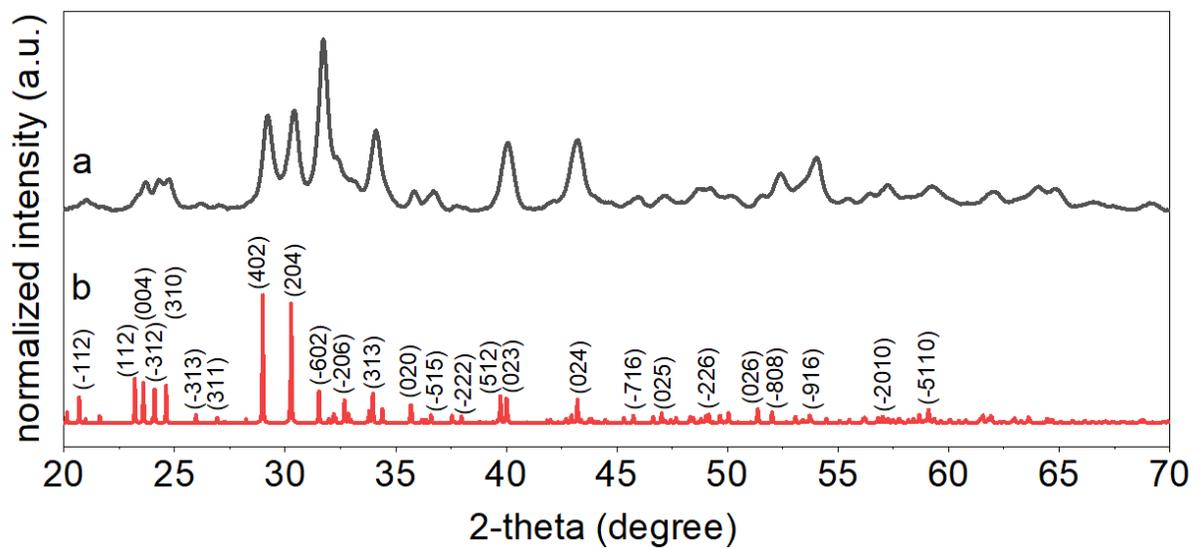

**Figure S5.** A semi-indexed XRD pattern of: (a) SOMS prepared by a solvothermal synthesis; and (b) a reported reference sample of $Na_2Nb_2O_6 \cdot H_2O$ (ICSD No. 55415). The major reflections associated with the products matched those of the reported reference sample.

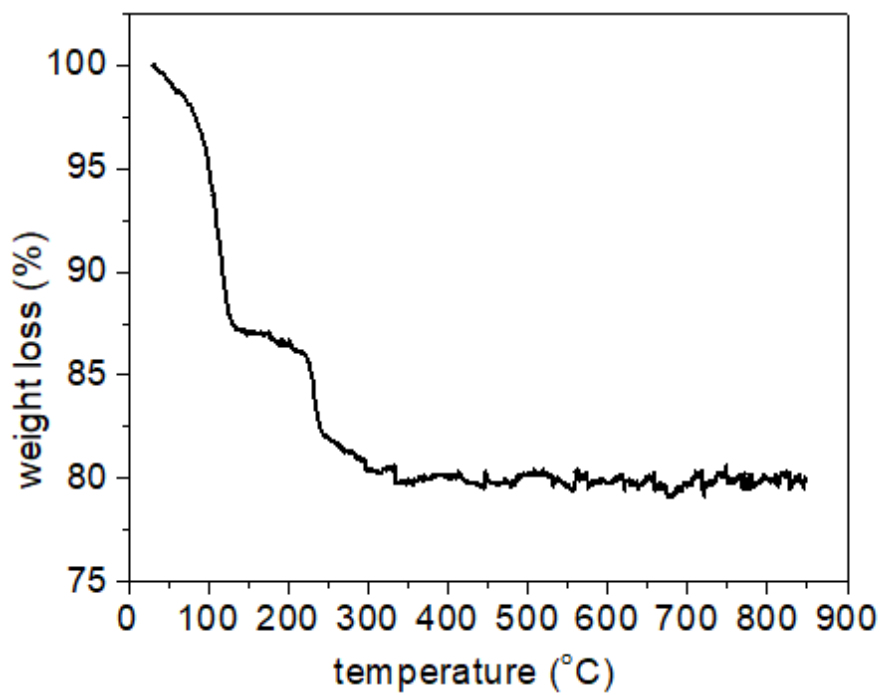

**Figure S6.** Thermally driven weight loss as obtained from a thermogravimetric analysis (TGA) of the SOMS products prepared by hydrothermal synthesis when heated at a rate of 1 °C/min from 30 to 850 °C under an ambient atmosphere. Following the loss of water from the surfaces and the lattice of the SOMS, these results indicate the relative thermal stability of the product.

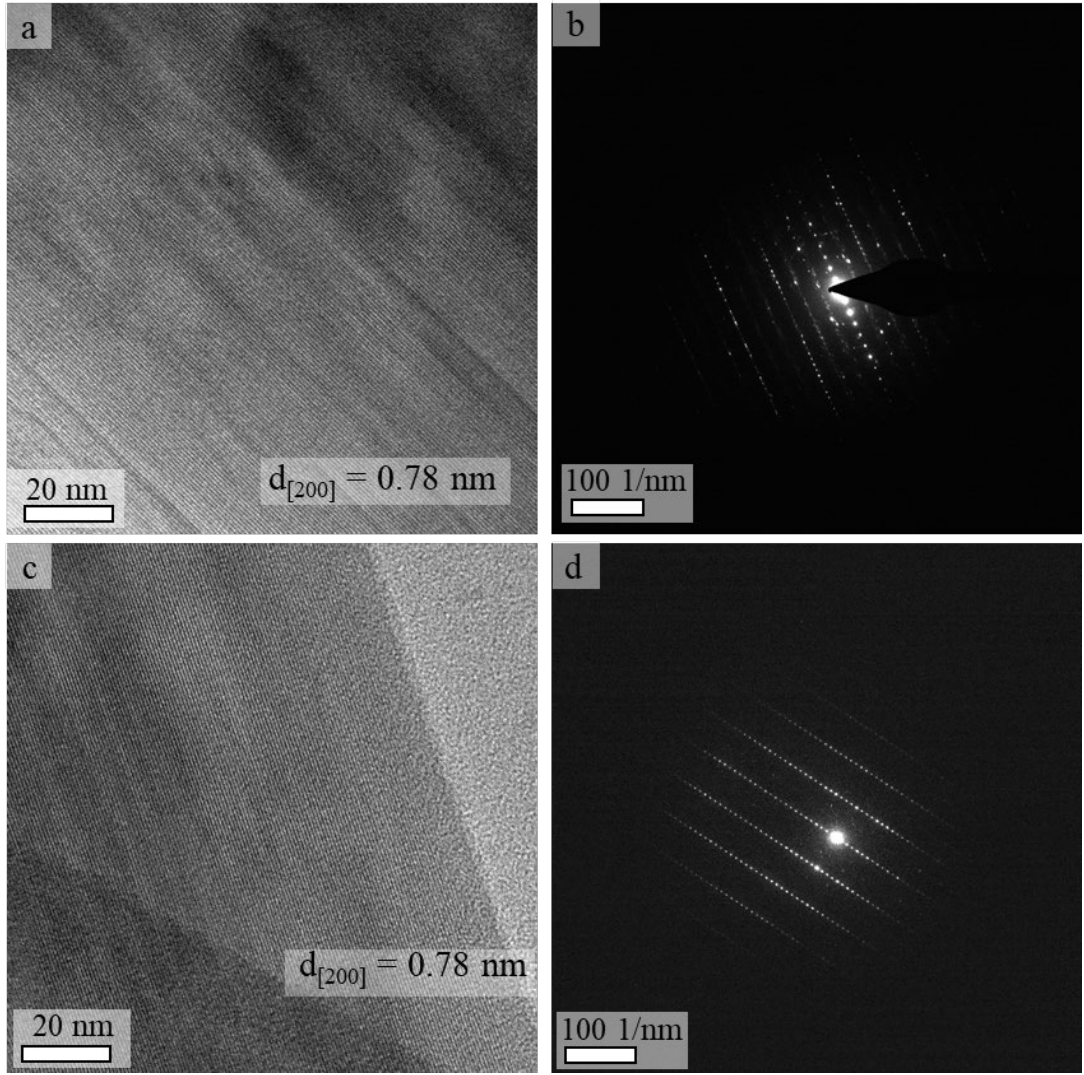

**Figure S7.** Aberration corrected HRTEM and SAED analyses at cryogenic temperatures of SOMS-based nanorods: (a,b) before neutron irradiation; and (c,d) after neutron irradiation. These analyses were performed with the assistance of the Facility for Electron Microscopy Research (FEMR) at McGill University using a Thermo Scientific Talos F200X G2 aberration corrected (S)TEM operated while holding the samples under cryogenic conditions (using a cryo-holder) to minimize potential damage or distortion to the samples from the incident focused electron beam.

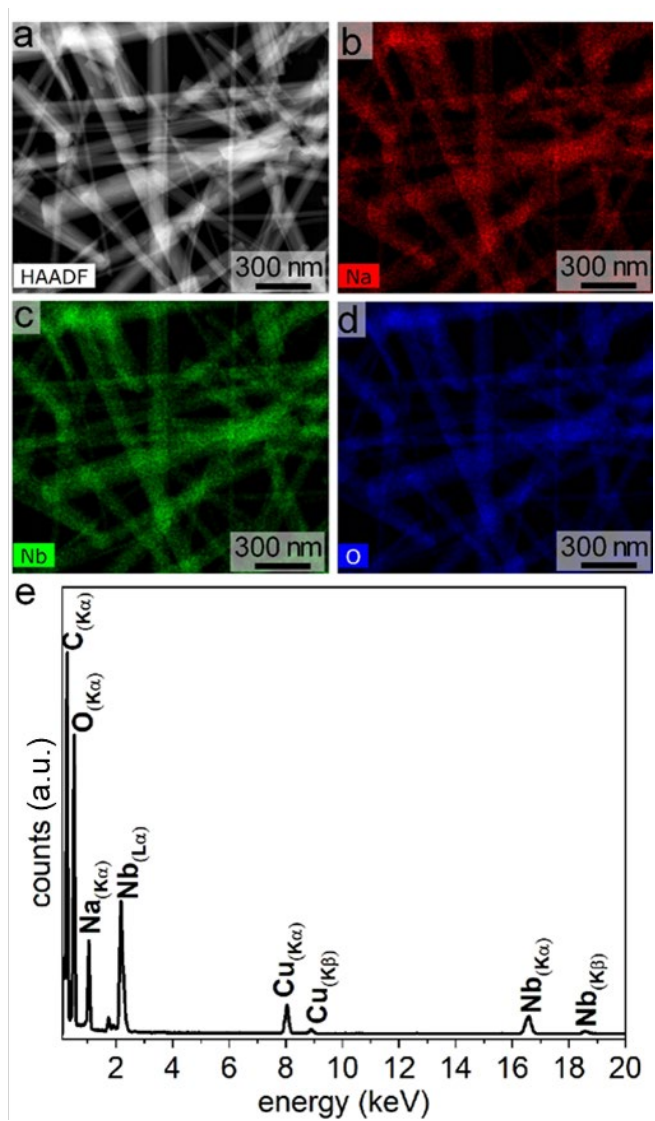

**Figure S8.** Additional results obtained by TEM and EDS analyses of SOMS-based nanorods after their exposure to energetic, 14.1 MeV neutrons. (a) A HAADF-based STEM image, and corresponding elemental maps obtained by EDS for (b) Na, (c) Nb, and (d) O. (e) An EDS spectrum corresponding to the nominal composition of these nanorods, which further confirmed the presence of Na, Nb and O in the product. The Cu signals in the spectrum originated from the Cu TEM grid used to support the sample during these measurements.

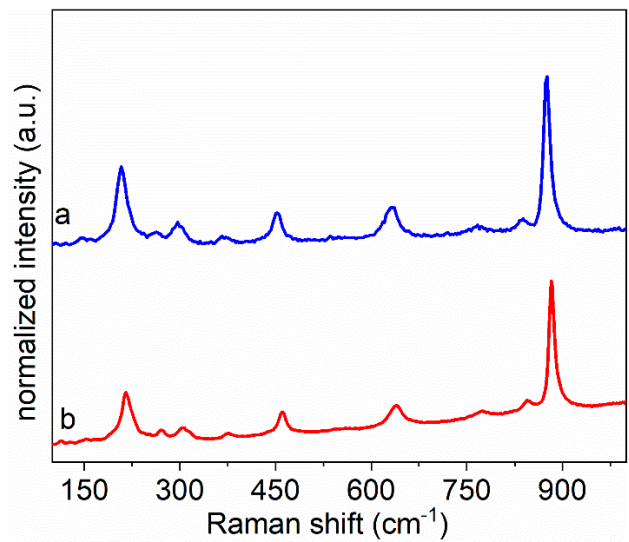

**Figure S9.** Room temperature Raman spectra obtained from the SOMS-based nanorods both (a) before and (b) after exposure to neutrons for 72 h.